\documentclass[preprint, amssymb, aps, prl]{revtex4-1}

\usepackage{amsmath}
\usepackage{graphicx}
\usepackage{epstopdf}

\DeclareMathVersion{mathchartertext}
\SetSymbolFont{letters}{mathchartertext}{OML}{mdbch}{m}{n}
\newcommand{\gchar}[1]{\mathversion{mathchartertext}$#1$\mathversion{normal}}
\DeclareMathVersion{mathchartertext}
\SetSymbolFont{letters}{mathchartertext}{OML}{mdbch}{m}{n}
\newcommand{\charmu}{\mathversion{mathchartertext}$\mu$\mathversion{normal}}
\newcommand{\um}{\charmu m\,}
\newcommand{\oneum}{1 \charmu m\,}

\def\abs#1{\mathopen| #1 \mathclose|}			% use instead of $|x|$ 
\begin{document}

\title{Tapering Enhanced Stimulated Superradiant Oscillator}

\author{J. Duris}
\affiliation{Department of Physics and Astronomy, UCLA, Los Angeles, California 90095, USA}
\affiliation{SLAC National Accelerator Laboratory, Menlo Park, CA}

\author{P. Musumeci}
\affiliation{Department of Physics and Astronomy, UCLA, Los Angeles, California 90095, USA}

\author{N. Sudar}
\affiliation{Department of Physics and Astronomy, UCLA, Los Angeles, California 90095, USA}

\author{A. Murokh}
\affiliation{RadiaBeam Technologies, Santa Monica, California, USA}

\author{A. Gover}
\affiliation{Faculty of Engineering, Department of Physical electronics, Tel-Aviv University, Tel-Aviv 69978 Israel}

\date{today}

\begin{abstract}
In this paper, we present a new kind of high power and high efficiency free-electron laser oscillator based on the application of the tapering enhanced stimulated superradiant amplification (TESSA) scheme. The main characteristic of the TESSA scheme is a high intensity seed pulse which provides high gradient beam deceleration and efficient energy extraction. In the oscillator configuration, the TESSA undulator is driven by a high repetition rate electron beam and embedded in an optical cavity. A beam-splitter is used for outcoupling a fraction of the amplified power and recirculate the remainder as the intense seed for the next electron beam pulse. The mirrors in the oscillator cavity refocus the seed at the undulator entrance and monochromatize the radiation. In this paper we discuss the optimization of the system for a technologically relevant example at 1 \gchar{\mu}m using a 1~MHz repetition rate electron linac starting with an externally injected igniter pulse.
\end{abstract}

\maketitle

%%%%%%%%%%%%%%%%%%%%
%%%%%%%%%%%%%%%%%%%%

%__________________________________________________________________________________________________________
%__________________________________________________________________________________________________________
\section{Introduction}

High wall-plug efficiency generation of intense, coherent, tunable electromagnetic radiation has many applications in all fields of science and technology. Free-electron based sources of radiation have been considered when the electromagnetic peak power is larger than the damage threshold for solid state materials or whenever it would not be possible to find a gain medium with a strong transition at the desired wavelength, as for example at the long (THz) and short (x-rays) wavelength ends of the electromagnetic spectrum.

For infrared, visible and ultraviolet region of the spectrum, laser amplifiers have recently demonstrated average power levels of kW, but with limited peak intensities \cite{laser_review}. Applications of high intensity and average power laser sources range from driving high gradient laser accelerators \cite{LWFA}, to inertial fusion or space-related applications \cite{IFRA}. At shorter wavelengths, UV lithography also requires high average power sources to increase the throughput of next generation integrated circuits \cite{EUV}.

Here we discuss a 100 kW-class average power light source which takes advantage of a novel mechanism for extracting energy from a relativistic electron beam, tapering enhanced stimulated super-radiant amplification \cite{Duris:NJP}. The energy exchange is mediated by the phase-synchronous ponderomotive potential wave experienced by a tightly bunched relativistic electron beam when propagating in a tapered magnetic undulator in the presence of a transverse electromagnetic wave, such as in a free-electron laser amplifier or inverse-free-electron-laser accelerator\cite{KMR}. The main advantages of this coupling scheme are the absence of nearby boundary or media (i.e. this is a vacuum plane-wave interaction), so that there are basically no mechanisms for the energy to flow out of the particle-field system. The key ingredients for reaching very high extraction efficiency are a high brightness electron beam, an intense drive laser to extract energy from the particles with high deceleration gradients, and a strongly tapered undulator with a prebunching section to maximize the amount of particles captured and decelerated in the ponderomotive potential trap.

Using an intense seed pulse, it has been recently experimentally demonstrated \cite{Sudar:Nocibur} that decelerating gradients of nearly 100 MV/m could be sustained for extended distances, enabling a large fraction of the energy in a relativistic electron beam to be extracted during passage in a short undulator magnet. The Nocibur experiment at the Accelerator Test Facility in BNL demonstrated over 30 $\%$ energy conversion efficiency at 10~$\mu$m wavelength in a single pass 54 cm long undulator and simulations indicate that comparable efficiency can be obtained at shorter wavelengths.

Taking advantage of superconducting radiofrequency techniques, modern electron accelerators can create high quality electron beams at high repetition rates with 100 kW to MW high average powers. Being able to extract a large fraction (up to 50 $\%$) of this power and convert it to electromagnetic radiation at any desired wavelength (tunability) has a potentially disrupting impact on many scientific and industrial fields. A fundamental issue to this approach is the availability of intense seed pulses at the desired wavelengths that can be used to extract the energy from the electron beams.

\begin{figure}[t]
\includegraphics*[width=120 mm]{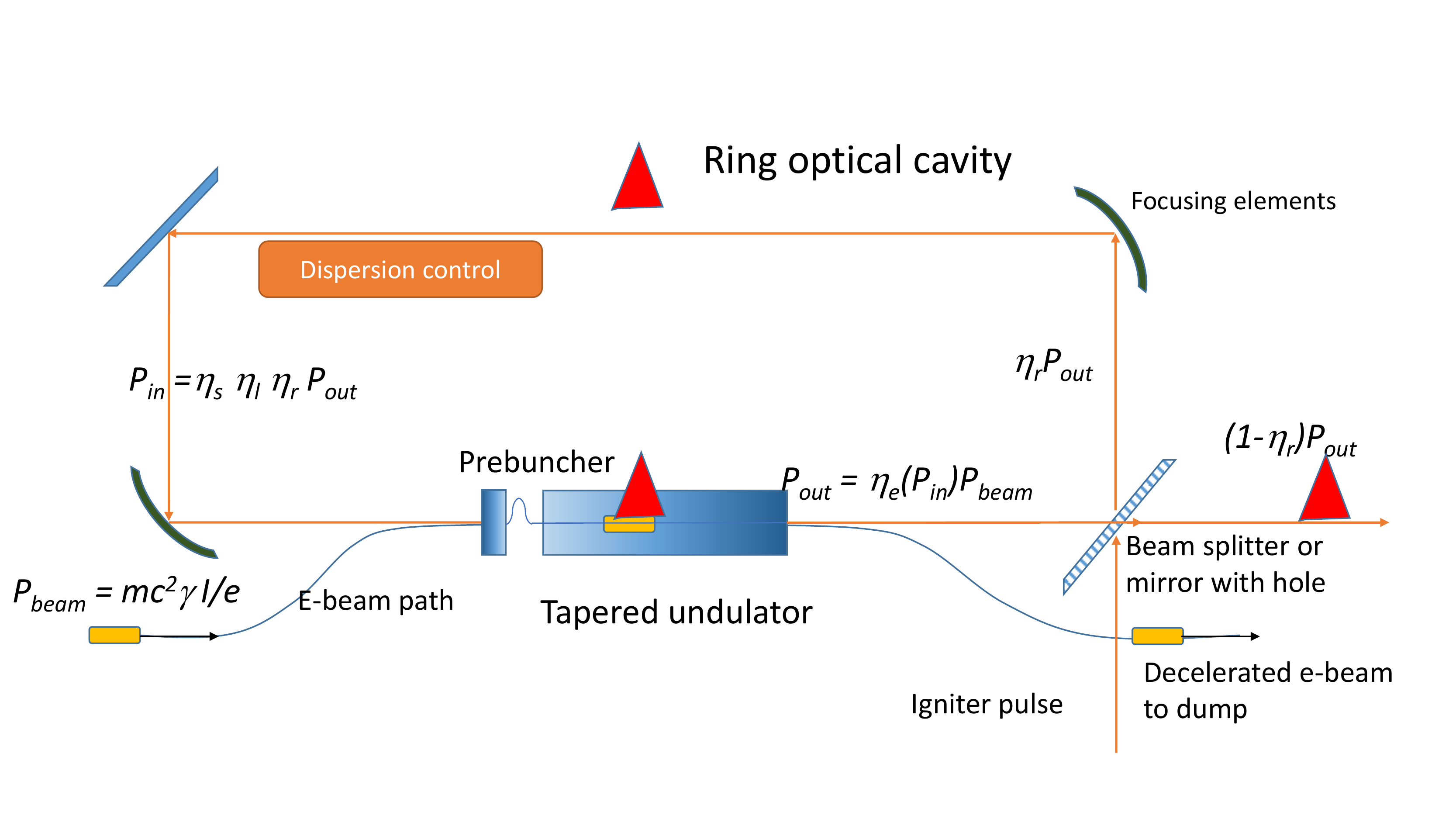}
\caption{Cartoon of the TESSO concept. A ring-like optical cavity is used to recirculate a small fraction of the electromagnetic radiation as a seed for the high efficiency decelerator. The cavity oscillation can be jump-started with an igniter pulse.}
\label{Fig:TESSO}
\end{figure}

The main idea behind this paper to address this problem is to marry the TESSA high efficiency energy extractor to an optical cavity and a high repetition rate electron beam. The output coupler of the cavity can be used to separate the TESSA output radiation from the electron beamline and redirect a fraction of the amplified power to the entrance of the undulator to serve as the intense seed for the deceleration of the next electron beam pulse and so on. A concept for this TESSA-based oscillator or TESSO is shown in Fig. \ref{Fig:TESSO}.

We envision two main application areas for TESSO. First, in regions of the electromagnetic spectrum --for example visible and IR-- where high peak power lasers are abundant, but lacking in high average power, and the proposed mechanism offers a path towards increasing by orders of magnitude the repetition rate (and the average power) of TW-class short pulse lasers.
Second, at shorter wavelengths --for example in the deep and extreme UV-- where TESSO might be the only viable solution for a high average power source. Note that in this latter case there will be no options for intense seed lasers, and one will need to develop solutions where the power builds up in the oscillator over several passes starting from a lower power seed.

The concept of a pre-bunched electron beam FEL oscillator, operating on the principle of stimulated-superradiance \cite{Gover:SSR}, was first proposed in \cite{Krongaus:2000,Gover:SSR2}. In these papers it was shown that the stable saturation point of a pre-bunched FEL oscillator provides high radiative energy extraction efficiency and potential high power operation, by taking advantage of most efficient phase-space dynamics of tight electron bunches, performing a synchrotron oscillation process in the ponderomotive potential well. It was recognized there, however, that arriving to this optimal steady state oscillation point, starting from superradiant emission of the bunches or a low power seed radiation, requires a particular strategy of beam energy temporal tapering during an oscillation build-up stage and overcoming a bistable state of the oscillator. Such a process would be harder to implement in the present case of interest of a tapered wiggler, because the small signal gain in the early stage of the oscillation build-up is suppressed.

In this article we avoid the problem of developing a strategy of oscillation build-up from low power, and concentrate on the analysis of a tapering enhanced stimulated superradiant oscillator – TESSO – in its optimal steady-state operation point, assuming that a single short pulse high power laser source is available in order to ignite the oscillator (see Fig. 1).

% https://en.wikipedia.org/wiki/Space-based_solar_power#Laser_power_beaming
% Laser power beaming[edit]
% Laser power beaming was envisioned by some at NASA as a stepping stone to further industrialization of space. In the 1980s, researchers at NASA worked on the potential use of lasers for space-to-space power beaming, focusing primarily on the development of a solar-powered laser. In 1989 it was suggested that power could also be usefully beamed by laser from Earth to space. In 1991 the SELENE project (SpacE Laser ENErgy) had begun, which included the study of laser power beaming for supplying power to a lunar base. The SELENE program was a two-year research effort, but the cost of taking the concept to operational status was too high, and the official project ended in 1993 before reaching a space-based demonstration.[49]
%In 1988 the use of an Earth-based laser to power an electric thruster for space propulsion was proposed by Grant Logan, with technical details worked out in 1989. He proposed using diamond solar cells operating at 600 degrees to convert ultraviolet laser light.
% More outlandish is the idea of interstellar travel

As a practical example we will examine the case of generating 1 $\mu$m radiation which is particularly interesting since it falls in a transparency window of the atmosphere. A high average power, micron wavelength light source facilitates interactions such as power beaming to high-bandwidth satellites, deorbit burning of space trash, boosting satellites to higher orbits\cite{benford1995space,PB,bennett1999fel}.%, and perhaps even small space craft launching \cite{laserablationpropulsion}. % TESSO can only do <MW which limits usefulness for small craft launching (people talk about 100MW for launches). That said, ablation propulsion could be useful in space to limit fuel, but focusing/tracking may be a challenge. I've removed refs to launching, but we could add again if you want.

%Later in this paper, we will turn to a design study for a high average power extreme ultra-violet (EUV) light source. Lithography is used in semiconductor electronics manufacturing and shorter wavelength light enables smaller feature manufacturing, which increases integrated circuit speeds and reduces power dissipation \cite{SemiconductorManufacturingOverview}. A lack of kW average power EUV light sources has halted integrated circuit feature size reduction recently \cite{MooresLawFailure}. Existing EUV light sources based on plasma discharge have been shown to be energetically inefficient and hard to scale to average powers required for semiconductor manufacturing \cite{PlasmaEUV}. Exploiting the high efficiencies and average powers achievable with superconducting rf accelerators via TESSO may enable kW power generation for EUV lithography.

The structure of the paper is organized as follows. We will first start from estimate of a single-pass module and cavity efficiency. Scaling laws are given so that the results of the paper can be easily adapted to other electron beam energies/radiation wavelengths. An experimental scenario with a 250 MeV beam will then be discussed in detail. We will then present the design of the optical cavity and perform numerical simulations to analyze the stability of the system to beam fluctuations. A new FEL simulation tool based on the well benchmarked Genesis code has been developed to follow the evolution of the field through multiple rountrips in the oscillator cavity. In the conclusions we outline the next steps towards the demonstration of a very high average power laser based on the TESSO principle.

%__________________________________________________________________________________________________________
%__________________________________________________________________________________________________________
\section{Efficiency estimate}

\subsection{Single pass}
In order to guide the choices in the TESSO oscillator design, we start by estimating the single pass efficiency of a TESSA amplifier for a helical undulator ignoring transverse effects and initially assuming that the gain in radiation intensity is small. In order to obtain rough analytical expressions for the efficiency, we will consider here the case in which only the magnetic field amplitude is tapered while the period is held constant throughout the interaction. We will also assume that the pre-bunching section driven by the seed laser beam effectively generates a tightly phased train of microbunches. These can then be injected at the proper phase in the decelerating tapered wiggler and trapped in the ponderomotive potential losing coherently most of their energy and contributing to the radiation gain.

The FEL equations of motion for a helical undulator are given by \cite{KMR}
\begin{eqnarray}
\frac{d\gamma}{dz}&=&-\frac{k K_l K}{\gamma} \sin{\psi} \\ \label{Eq:EnergyEvo}
\frac{d\psi}{dz}&=&k_w \left( 1-\frac{k (1+K^2)}{2 k_w \gamma^2} \right) \label{Eq:PhaseEvo}
\end{eqnarray}
where $K_l$ for a circularly polarized Gaussian laser is related to the laser power $P$ and the $1/e^2$ laser waist $w$ as $K_l = \frac{e_0}{k m c^2} \sqrt{\frac{Z_0 P}{\pi w^2}}$ with $Z_0$ is the vacuum impedance.

Strong coupling and resonant interaction is obtained when the relative phase of the electrons in the ponderomotive potential wave $\psi$ is stationary in Equation~\ref{Eq:PhaseEvo}, yielding the condition for the resonant energy
\begin{equation}
\gamma_r=\sqrt{\frac{k}{2k_w}(1+K^2)}
\label{Eq:ResEnergy}
\end{equation}

In order to maintain resonance as the electron beam is decelerated, the undulator parameter must be varied so that the resonant energy changes consistently with the available ponderomotive decelerating gradient.  This is achieved by equating Eq.~\ref{Eq:EnergyEvo} with the derivative of Eq.~\ref{Eq:ResEnergy} to obtain, assuming constant period, an expression for the variation in the undulator strength parameter $dK/dz$
\begin{equation}
\frac{dK}{dz} = - 2 k_w K_l \sin \psi_r
\end{equation}
Integration of this expression with the assumption that the product $K_l \sin{\psi_r}$ remains approximately constant along the undulator yields a linear variation of $K$ over the undulator length $L_w=\lambda_w N_w$ with $K(z = L_w) - K_0 = \Delta K = -4\pi K_l \sin \psi_r N_w$.

For a fully prebunched electron beam where particles are trapped in the ponderomotive potential to maintain resonant interaction along the entire undulator length, the energy extraction efficiency can then be approximated as
\begin{equation}
\eta_e=1-\frac{\gamma_f}{\gamma_0} \cong \frac{K_0 \abs{\Delta K}}{1+K_0^2}
\label{Eq:ExtractionEfficiency}
\end{equation}
where $\gamma_{0,f}$ are the resonant energies at the entrance and exit of the undulator respectively, $K_0$ is the initial undulator parameter, and the approximation holds in the limit $\Delta K << K_0$.

\begin{figure}[t]
\includegraphics[width=0.45\textwidth]{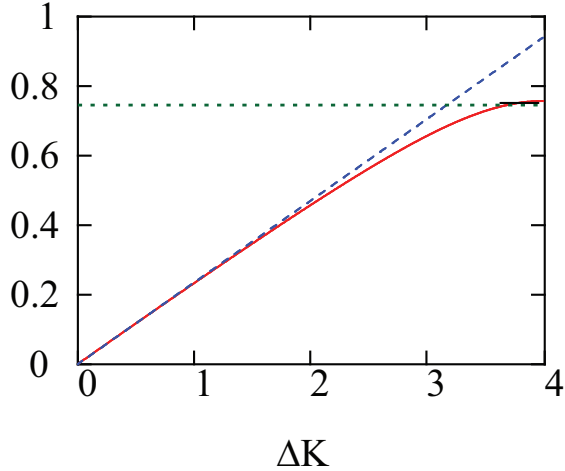}
\caption{Efficiency for TESSA single pass energy extraction from a relativistic electron beam in the limit of low radiation gain for $K_0$=4 (red curve). The approximation in Eq. \ref{Eq:ExtractionEfficiency} is also shown (blue dashed). The maximum efficiency limit $\eta_{e,m}$ is indicated by a dotted green line.}
\label{Fig:efficiency}
\end{figure}

The efficiency increases with the number of undulator periods $N_w$ up to the maximum limit given by $\eta_{e,m} = 1-\sqrt{1+K_0^2}^{-1} \approx 1 - 1/K_0$ for a large initial undulator parameter (i.e. $K_0 >> 1$).  This limit corresponds to a final undulator parameter equal to zero (i.e. $\Delta K = K_0$). In practice though, the electrons would dephase towards the end because the trap amplitude is proportional to $\sqrt{K}$, and therefore $K(z=L_w)$ should be at least larger than $K_0$/4. A plot of $\eta_e$ and the approximation in Eq.\ref{Eq:ExtractionEfficiency} is shown in Fig. \ref{Fig:efficiency}.

The actual efficiency will depend on how many electrons are captured and decelerated in the ponderomotive potential. A choice of resonant phase near $\pi/2$ maximizes the decelerating gradient at the expense of the number of particles captured, whereas a resonant phase near zero optimizes beam capture with little deceleration. A resonant phase near $\pi/4$ offers a compromise between accelerator acceptance and decelerating gradient, optimizing the energy extraction.

Diffraction is responsible for variation of electromagnetic intensity (and therefore $K_l$) along the undulator. Including a simple model based on an Hermite-Gaussian mode propagating in free space \cite{Siegman:lasers}, it is possible to calculate that for an undulator length with a linear $K$ taper, the maximum energy change is obtained when the laser is focused to minimum waist position at $z_w=L_w/2$ with a Rayleigh range of $z_r=3L_w/20$ \cite{Duris:dissertation}. Averaging $K_l$ over the length of the undulator yields $\frac{3}{10} \sinh^{-1}(\frac{10}{3}) = 0.576$ times the peak value of the laser normalized vector potential $K_{l,peak}$.

If the undulator is designed to accommodate an increasing radiation power (high gain TESSA regime), Eq. ~\ref{Eq:ExtractionEfficiency} can be considered a lower bound to the single pass extraction efficiency since the additional radiation field may be used to decelerate the beam further and the efficiency can be higher.

Invoking energy conservation, the radiation power gained at the exit of the TESSA amplifier can then be written as $\Delta P= \eta_e(P_{in}) P_b$ where $P_b$ is the beam power and
\begin{equation}
\eta_e(P_{in}, K_0, N_w) \cong \frac{K_0}{1+K_0^2} 0.576 \cdot 4\pi \frac{e_0\sqrt{Z_0 P_{in}}}{k m_0 c^2} \frac{1}{\sqrt{0.15 \lambda \lambda_w}} \sin \psi_r \sqrt{N_w}
%\eta_e(P_{in}, K_0, N_w) \cong \frac{K_0}{1+K_0^2} 0.576 \cdot 4\pi \frac{e_0\sqrt{2 Z_0 P_{in}}}{k m_0 c^2} \frac{1}{\sqrt{0.15 \lambda \lambda_w}} \sin \psi_r \sqrt{N_w}
\end{equation}
which in engineering units can be simplified to
\begin{equation}
\eta_e(P_{in}, K_0, N_w) \cong \sqrt{\frac{K_0^2}{1+K_0^2}} \frac{3.56}{\gamma_0} \sqrt{P_{in}(GW)*N_w}
%\eta_e(P_{in}, K_0, N_w) \cong \sqrt{\frac{K_0^2}{1+K_0^2}} \frac{2.5}{\gamma_0} \sqrt{P_{in}(GW)*N_w}
\label{singlepass_efficiency}
\end{equation}

\begin{figure}
\includegraphics[width=0.9\textwidth]{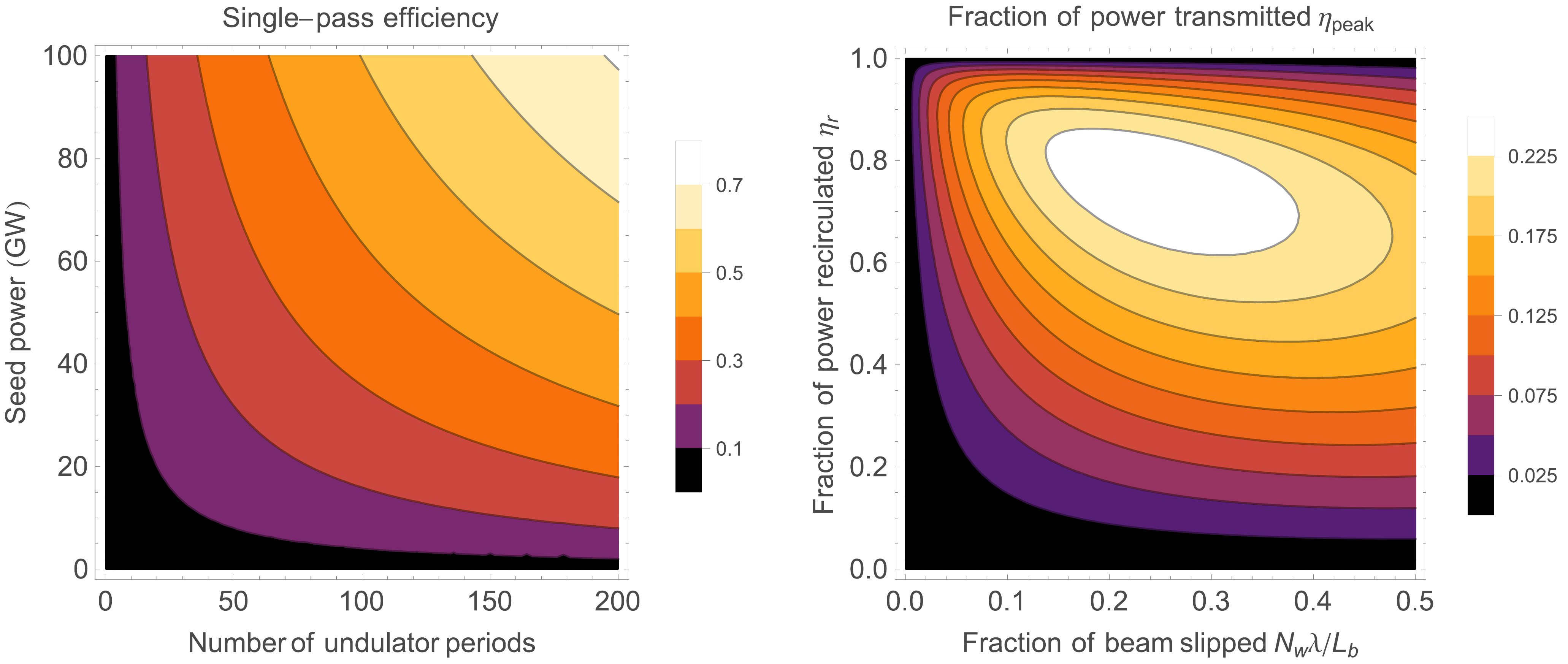}
\caption{Left: Single pass TESSA power extraction efficiency $\eta_e$ for a 250 MeV beam, $K_0$ = 4 as a function of input power and number of undulator periods. Right: Oscillator efficiency as a function of number of periods and output coupler reflectivity.}
\label{Fig:etas}
\end{figure}

A plot of Eq. \ref{singlepass_efficiency} as a function of input power and number of undulator periods for a 250 MeV beam, $K_0$ = 4 is shown in Fig. \ref{Fig:etas}a.

\subsection{Oscillator efficiency}
In order to obtain an expression for the TESSO cavity efficiency, we need to include the losses on the cavity round trip $\eta_l$, the output coupler reflectivity $\eta_r$ and consider the slippage effects.

In the undulator, the radiation slips forward relative to the electron beam by one wavelength per undulator period for a total slipped length of $\lambda N_w$. When the undulator is tapered to take advantage of the increasing radiation power at the head of the electron beam, radiation overlapping the tail of the electron beam may be insufficient to maintain acceleration for electrons there, and as a result, the length of the radiation pulse region with constant power is $L_b-\lambda N_w$. This constant power region of the radiation must be stretched during a cavity roundtrip to cover the entire length of the electron bunch on the next pass, reducing the peak power by a fraction $\eta_s = 1 - \lambda N_w / L_b$ where we neglected the energy contained in the tail of the radiation temporal profile.

Consequently, the steady state power at the entrance of the undulator available for decelerating the electron beam can be written as
\begin{equation}
P_{in,ss} = \eta_s \eta_r \eta_l P_{out} = \eta_s \eta_r \eta_l (P_{in,ss}+\eta_e(P_{in,ss}) P_b)
\label{Eq:balance_equation}
\end{equation}
which represents the fundamental power balance equation for the oscillator cavity that allows us to find the steady-state of the system.

Since the efficiency $\eta_e$ depends on the input power, in order to find the steady state intracavity recirculating power, we combine Eq. \ref{Eq:balance_equation} with Eq. ~\ref{Eq:ExtractionEfficiency} and solve for $P_{in,ss}$ yielding:
\begin{equation}
P_{in,ss}(GW) = \frac{K_0^2}{1+K_0^2}\frac{12.7 N_w}{\gamma_0^2} P_b(GW)^2
\frac{\eta_s^2 \eta_r^2 \eta_l^2}{\left(1-\eta_s \eta_r \eta_l\right)^2}
\label{Eq:recirculating power}
\end{equation}

The optimal configuration is found by maximizing the output coupled oscillator power which is given by $P_{osc}=P_{out} (1-\eta_r)$ as a function of undulator length ($N_w$) and output coupler reflectivity ($\eta_r$). The result of the optimization depends on the cavity losses and is shown in Fig. \ref{Fig:efficiency} for $\eta_l = 0.8$. In this case the optimal values are $N_w = L_b/3 \lambda$ and $\eta_r$ = 0.66.

With these choices, the output oscillator pulse peak power to beam power ratio is $\eta_{power}=P_{osc}/P_{beam}=(1-\eta_r)P_{out}/P_{beam}$. The output energy per pass is obtained multiplying the peak power by the length of the radiation pulse ($\sim \eta_s L_b$), obtaining for the energy extraction efficiency
\begin{equation}
\eta_{energy} \cong \frac{K_0^2}{1+K_0^2}\frac{L_b}{\lambda \gamma_0^2} \eta_l P_b(GW)
\end{equation}
The outcoupled oscillator energy per pass is $U_{osc} = \eta_{avg} U_b$ where $U_b$ is the total energy of the electron beam.

% Need to match the results of this section with what comes later
% numerical prefactor is 0.8 or about unity:  s(1-r){12.7,6.25}(1-s) (s r l)/(1-s r l)^2/.{l->0.8,r->2/3,s->2/3}

Dividing the output radiation energy by the photon energy  $h c/ \lambda$ we obtain a simple expression for the number of photons that can extracted from the oscillator per pass
\begin{equation}
N_\gamma \approx \frac{K_0^2}{1+K_0^2}\alpha N_b^2 % numerical prefactor is 1.2 or about unity: 0.697*2s(1-r){12.7,6.25}(1-s) (s r l)/(1-s r l)^2/.{l->0.8,r->2/3,s->2/3}
\label{Eq:PhotonsTransmitted}
\end{equation}
where $\alpha$ is the fine structure constant, and $N_b$ is the number of electrons in the bunch. The number of photons is independent on the radiation wavelength or the electron energy and just goes as the square of the number of electrons participating to the interaction. This scaling is characteristic of superradiant emission \cite{Gover:SSR} and is a direct consequence of the fact that the TESSO cavity ensures that electrons in one bunch generate the radiation field stimulating the electrons in the subsequent bunch to emit coherently. The average oscillator power is proportional to the electron beam repetition rate.

Equation~\ref{Eq:PhotonsTransmitted} should be considered as a rough estimate as the analysis ignored the possibility of varying the undulator period, which has been found to increase power extraction efficiency as seen in \cite{Duris:NJP} and the additional benefits due to the steeper taper to take advantage of the additional stimulated radiation. At the same time, a rough account of diffraction is included but we ignored many three dimensional effects (beam emittance, transverse radiation mode, etc.) and assumed trapping of a fully bunched electron beam. We will consider all of these effects with the help of self-consistent particle tracking simulations in the next section. Still, these estimates are a reasonable starting point to guide the design of the oscillator.

Another thing to note about Eq.~\ref{Eq:PhotonsTransmitted} is that the number of photons transmitted is independent of the current, suggesting optimal operation with very long electron bunches (having many particles per bunch). On the other hand, the analysis above shows that an undulator length of $L_w=\lambda_w N_w=\lambda_w L_b/3 \lambda$ maximize performance. For a long electron beam, the undulator might become so long that the seed radiation size $w_0=\sqrt{z_r\lambda/\pi}=\sqrt{0.15 L_w\lambda/\pi}$ will become comparable to the undulator gap. In practice, the constraints imposed by a particular undulator technology choice will limit the optimal length of the electron bunches for this application.

%__________________________________________________________________________________________________________
%__________________________________________________________________________________________________________
\section{TESSO oscillator design}

%Choice of parameters
\subsection{TESSO parameters}

To illustrate better the formulae derived in the previous session, let us considered a practical example of a \oneum\ oscillator based on the TESSO concept. Ideally, the electron beam energy should be made small to reduce linac cost and size, but large enough to achieve large currents without compromising the beam brightness via compression. This is crucial to shorten the undulator length which is directly proportional to $L_b$. We start with a 1~nC, 250~MeV electron beam compressed to a 2~ps duration for a current of 500~A, and peak beam power of 125~GW. The parameters for this example and the simulations of the rest of the paper are shown in Table \ref{Tab:Params}.

For a 2 ps long electron bunch, the undulator length to optimize energy extraction is $\sim$ 4~m (i.e. $N_w \cong L_b/3 \lambda$). We can then calculate the seed laser focal parameters to optimize the energy exchange between electron beam and radiation. In the case where the seed dominates the interaction, the laser waist should be placed at the center of the undulator with a Rayleigh range about 15\% the length of the undulator. Using a numerical optimization model which takes into account for the growth of the stimulated radiation, it is found that the region of peak intensity shifts downstream of the seed waist. Therefore we choose a waist at $z$ = 1.6~m and Raleigh range of 45~cm to maximize the electromagnetic field intensity along the undulator and optimize the electron beam deceleration.

\begin{table}[b]
\caption{Decelerator parameters}\label{Tab:Params}
\begin{ruledtabular}
\begin{tabular}{lc}
Electron beam & \\
\hline
Input energy & 250 MeV \\
Emittance & 1 mm-mrad \\
Transverse size (rms) & 40 \um \\
Bunch length $L_b$ FWHM &  2ps \\
Energy spread & 0.1 $\%$ \\
Peak current & 500 Amp \\
Charge & 1 nC \\
Repetition rate & 1~MHz \\
Beam power & 125 GW (peak); 250 kW (average)\\
\hline
Seed laser & \\
\hline
Wavelength & 1 \um \\
Power & 50 GW \\
Focal spot size (1/e$^2$) & 350 \um \\
Rayleigh range & 45 cm \\
Cavity length & 300~m \\
\hline
Undulator & \\
\hline
Geometry & helical Halbach array \\
Residual magnetization & 1.42~T \\
Initial period length & 2.7 cm \\
Initial K value & 4.2 \\
Length & 4 m \\
Gap & $>$ 3 mm \\
\end{tabular}
\end{ruledtabular}
\end{table}

The oscillator average power may be increased adjusting the linac repetition rate. In our example we consider 1 MHz which is the repetition rate of superconduncting RF linacs for future light sources \cite{LCLS2} which would require a fairly long 300 m cavity. Increasing the linac repetition rate to 5 MHz will shorten the cavity length to 60 m and proportionally increase the output power of the oscillator. Nevertheless, much higher repetition rates may be undesirable as the cavity roundtrip for pulse synchronization becomes too short. As we will see below, a relatively large distance between the undulator and the cavity mirrors is important to take advantage of diffraction to keep the radiation intensity below the damage threshold on the optics.

%_______________________________________________
\subsection{Prebunching}

A prebunched beam is needed to match the electron beam longitudinal phase space to the acceptance area of the decelerator. Prebunching may be achieved by imparting an energy modulation onto the beam with a short undulator section, and then converting that energy modulation into a density modulation by applying $R_{56}$ via a drift or magnetic chicane. A variable strength chicane enables fine adjustment of the phase delay for locking the prebunched beam with the ponderomotive bucket. Prebunching with an energy modulator and chicane enables bunching factors approaching 0.6 and may help entrain $\sim 60\%$ of the electrons into the ponderomotive wave \cite{STELLA2}. Cascading prebunchers has been suggested to increase to $>$ 90 $\%$ the captured fraction by concentrating the phase space further \cite{Sudar:doublebuncher}.

\begin{figure}[t]
\includegraphics[width=120 mm]{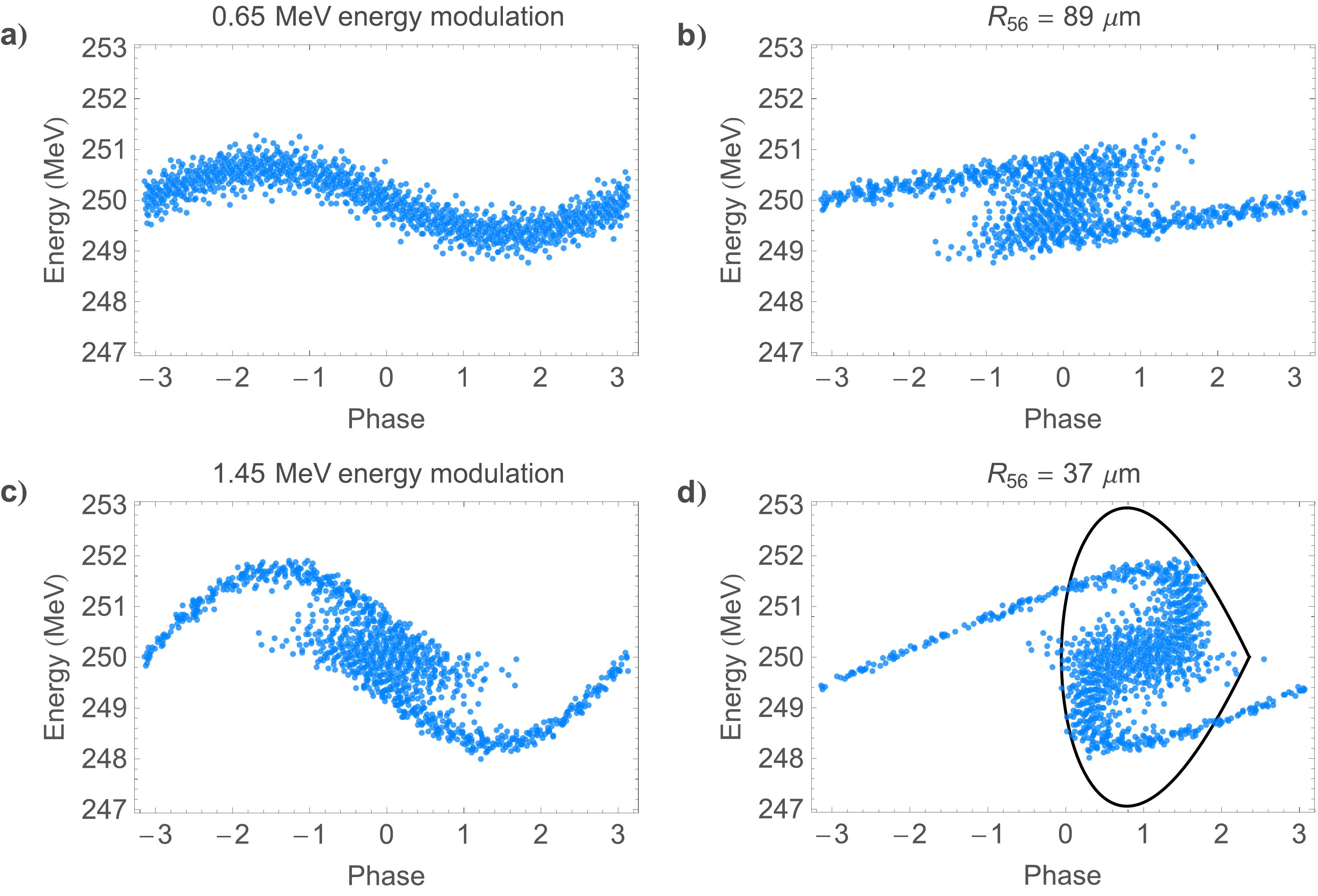}
\caption{Prebunching via cascaded energy modulator and chicane sections.}
\label{Fig:Prebunching}
\end{figure}

% sims use 89um and 37um R56
In order to maximize the efficiency we adopt the two stage prebuncher for this design study. The phase space evolution is shown in Figure~\ref{Fig:Prebunching}. A single period undulator section ($\lambda_w$=55 mm) is designed to impart a 650~keV amplitude sinusoidal energy modulation with the 50~GW seed laser power, and this energy modulation is slightly over compressed with a chicane with R$_{56}$=~143~$\mu$m. The bunched beam is then injected into a subsequent 4-period long energy modulator ($\lambda_w$=30~mm), imparting a 1.45~MeV amplitude modulation to the bunched beam. A final chicane (R$_{56}$~=~40~$\mu$m) serves the purpose of compressing the beam to maximize the fraction trapped within the ponderomotive bucket in the subsequent decelerator section.

It should be noted that the large $R_{56}$ required by the bunchers will add an additional delay between the radation and the electron beam in the cavity. This must be compensated by adding a corresponding optical path length in the radiation path for example using extra cavity mirrors.

%_______________________________________________
\subsection{Undulator taper design}

\begin{figure}[t]
\includegraphics[width=160 mm]{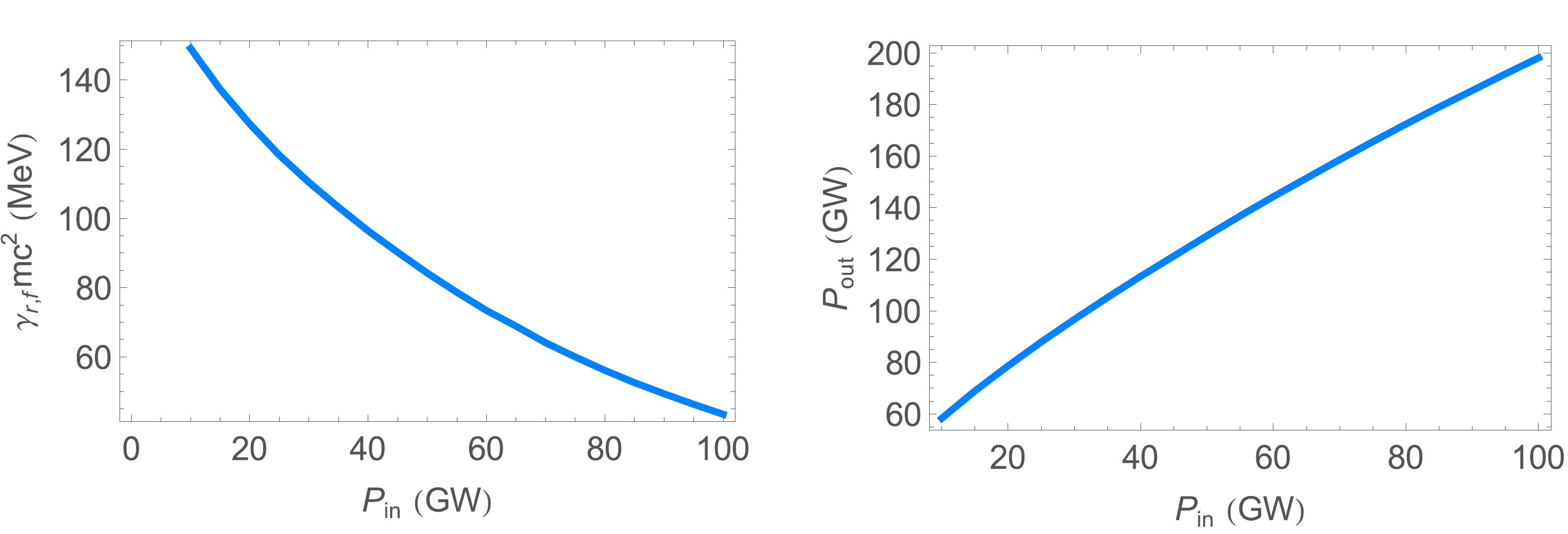}
\caption{Output electron beam energy, output power for undulators optimized for a given input power.}
\label{Fig:GITS_power_scan}
\end{figure}

We used the Genesis informed tapering scheme (GITS) to calculate the optimal taper of both magnetic field amplitude and undulator period for a range of initial seed laser powers, starting from the parameters in Table \ref{Tab:Params}. GITS has been described in detail in \cite{Duris:NJP} and is an optimization algorithm based on the well benchmarked 3D FEL code Genesis \cite{Genesis} to calculate the optimal variation of the undulator parameters taking into account the growth of the radiation along the undulator. Figure~\ref{Fig:GITS_power_scan} displays the output electron beam resonant energy and laser power for undulators optimally designed for a given input seed laser power.

\begin{figure}[t]
\includegraphics[width=140 mm]{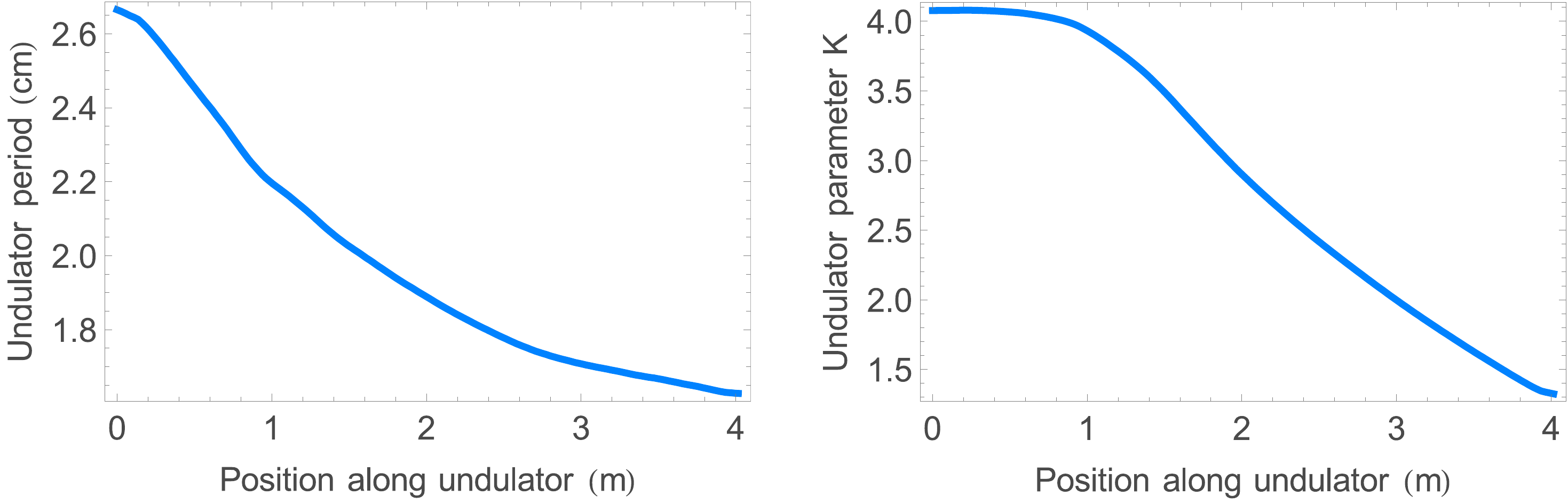}
\caption{Undulator period (left) and normalized undulator amplitude (right) as a function of undulator length for the reference 50 GW input seed power case.}
\label{Fig:GITS_period_and_amplitude}
\end{figure}

For more realistic results, instead of keeping a constant undulator period and varying only the magnetic field amplitude, the Halbach permanent magnet undulator builder equation \cite{Halbach:Undulators} is used to obtain the magnetic field amplitude for a given undulator period and gap. NdFeB magnets with 1.42 T residual magnetization are used. The undulator gap is varied so that $>$99.5\% of the radiation power clears the magnets, resulting in a minimum gap of 3~mm. The resulting undulator period and magnetic field taper for the reference 50 GW seed power are shown in Fig. \ref{Fig:GITS_period_and_amplitude}.

\begin{figure}[t]
\includegraphics[width=80 mm]{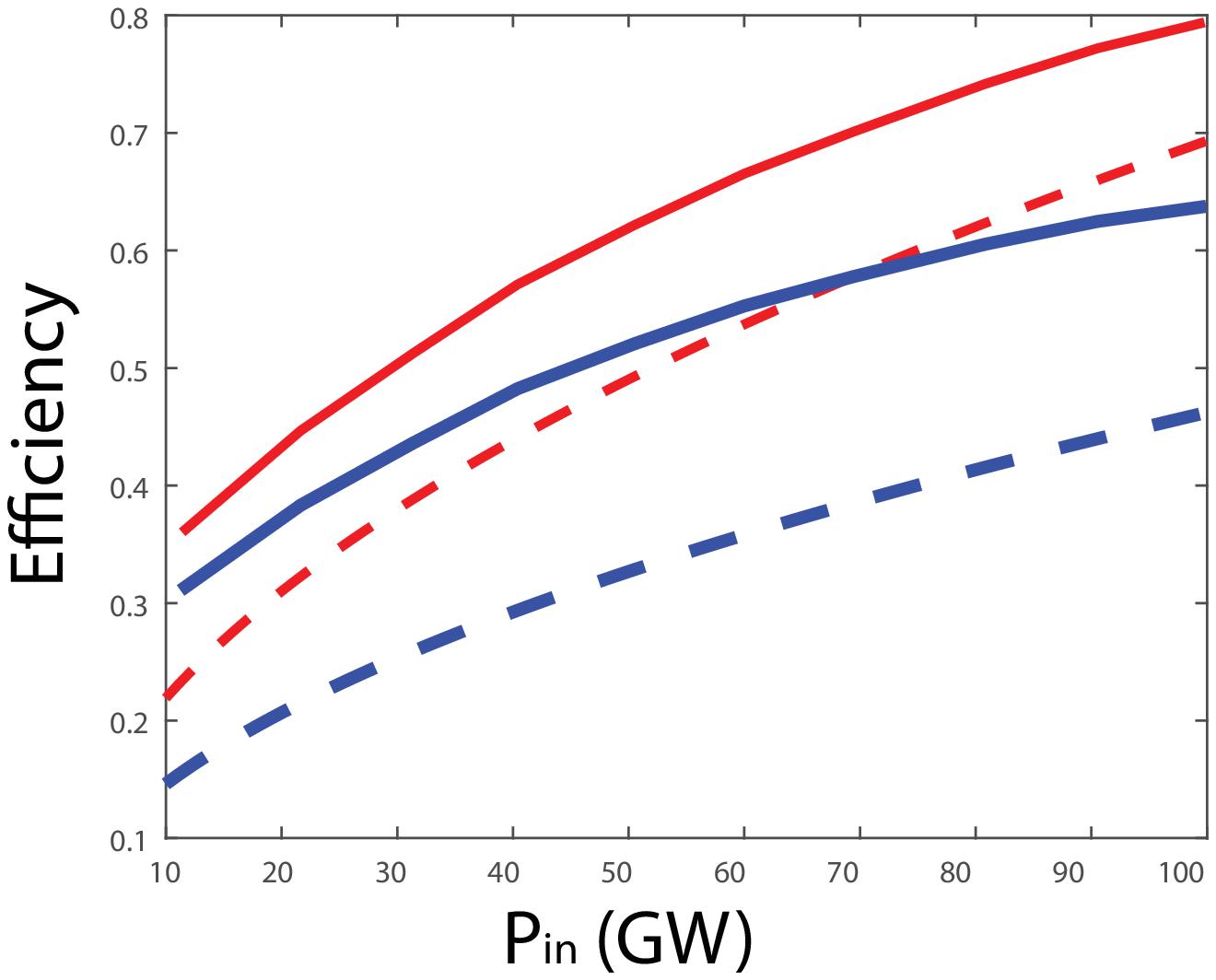}
\includegraphics[width=80 mm]{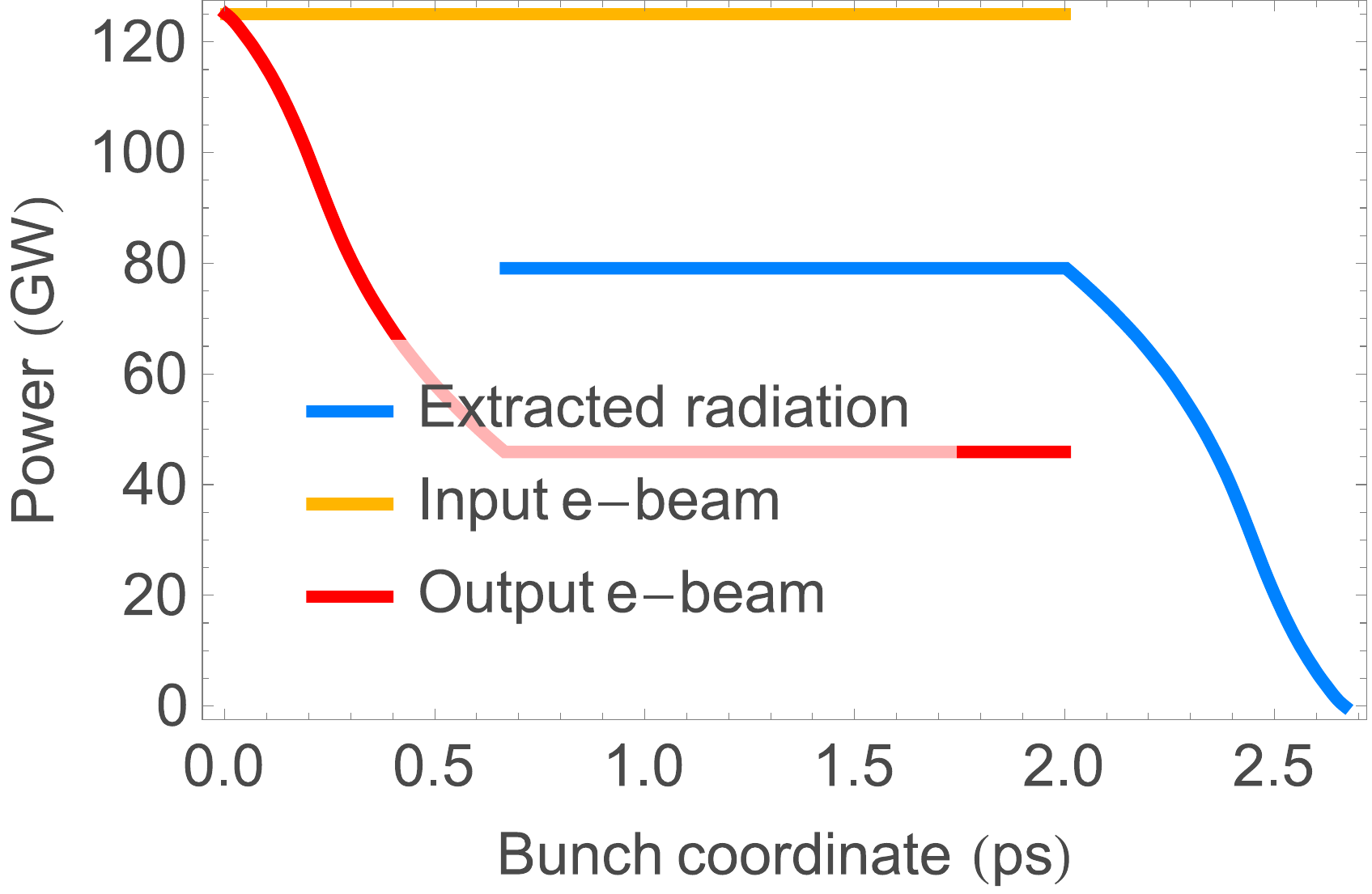}
\caption{a) Peak power (blue) and energy (red) efficiency from the GITS simulation results. The theoretical estimates $\eta_e$ and $\eta_e \cdot \eta_s$ for a constant period case with $N_w=200$ are also shown for reference. b) Time-domain view of the beam and radiation power at the entrance and exit of the undulator.}
\label{Fig:GITS_efficiency}
\end{figure}

Fig.~\ref{Fig:GITS_efficiency} compares the efficiency estimates of the previous section with the actual efficiency retrieved from the Genesis simulations. Whereas the theoretical estimates assumed no gain in the decelerating field, the significantly improved results from the numerical optimization are easily explained due to the fact that GITS designs the taper taking into account the extra field available from the amplification of the radiation resulting in a higher decelerating gradient.

An analysis of this simulation case provides a clear example of the effect of the slippage on the oscillator output which is the reason of the difference between peak power and energy transfer efficiency. The electron beam is 600 wavelengths long (flat-top, 2~ps long beam current profile), and the input seed radiation temporal profile is assumed to be initially perfectly synchronized with the e-beam and have exactly the same duration. Due to the FEL resonance condition, the number of radiation wavelengths slipped is equal to the number of undulator periods which in the simulations increases more or less linearly from $N_w$=160 periods at 10~GW input power to 240 periods at 100~GW seed power for the fixed 4~m undulator length.

The longitudinal phase space at the undulator exit is a low energy beam with a very strong positively chirped tail (shown in Fig. \ref{Fig:GITS_efficiency}b). Since the steady state fraction of laser is shorter than the input electron beam, the recirculated pulse must be stretched to cover the entire electron beam for the next pass. This can be accomplished by spectrally filtering or introducing dispersion on the recirculated radiation. The efficiency reduction associated with this manipulation was described by $\eta_s$ in the previous section. The stretching should also account for expected electron-laser time of arrival jitter in order to ensure reliable operation for hundreds or thousands of passes of the oscillator.

%__________________________________________________________________________________________________________
%__________________________________________________________________________________________________________
\subsection{Cavity design}

The optical cavity round-trip length is set by the 1~MHz electron bunch repetition rate to be $L_{cav}$=300~m. In principle, it may be possible to make multiple passes through the cavity with the benefit of dividing the resulting cavity length by the number of passes to reduce the overall size of the oscillator. This approach has many disadvantages though including higher optical losses and higher thermal load on the mirrors. On the other hand if low-loss mirrors are available, the cavity can be folded reducing considerably the footprint of the overall system.

A simple ring cavity design using a stable resonator to provide $z_r$~=~45 cm near the center of the 4~m undulator can be designed using the near concentric optical resonator formulas discussed in Siegman \cite{Siegman:lasers}, assuming no losses and no optical gain from the electron beam with the undulator as a simple drift. The radii of curvature of the mirrors are slightly larger than half length of the cavity i.e. $R = L_{cav}/2 + \Delta L$ so that the spot size at the waist and at the mirrors can be written
\begin{eqnarray}
w_0^2 = \frac{L_{cav}\lambda}{\pi} \sqrt \frac{\Delta L}{4 L_{cav}} \\
w_{mirror}^2 = \frac{L_{cav}\lambda}{\pi} \sqrt \frac{4 L_{cav}}{\Delta L}
\end{eqnarray}

In order to achieve $z_r = 45$ cm, we have $\Delta L = 2.7$ mm and a waist size at the mirrors of 25~cm. This large spot size helps in terms of controlling the power load on the mirrors. With $~$ 100 kW recirculating power in the cavity, the incident power on the mirrors is $<$ 1 kW /cm$^2$ so that it should be possible to water cool the substrate to dissipate the excess heat.

In order to maintain a few percent level, energy gradient stability with this design assuming mirrors with linear coefficient of thermal expansion of up to 10$^{-5}$ per degree typical of glass or metal mirrors, temperature stability of a few tenths of a degree Celsius is required. A shorter cavity length relaxes this requirement but increases the intensity on the mirrors, necessitating more cooling.

\begin{figure}[t]
\includegraphics[width=160 mm]{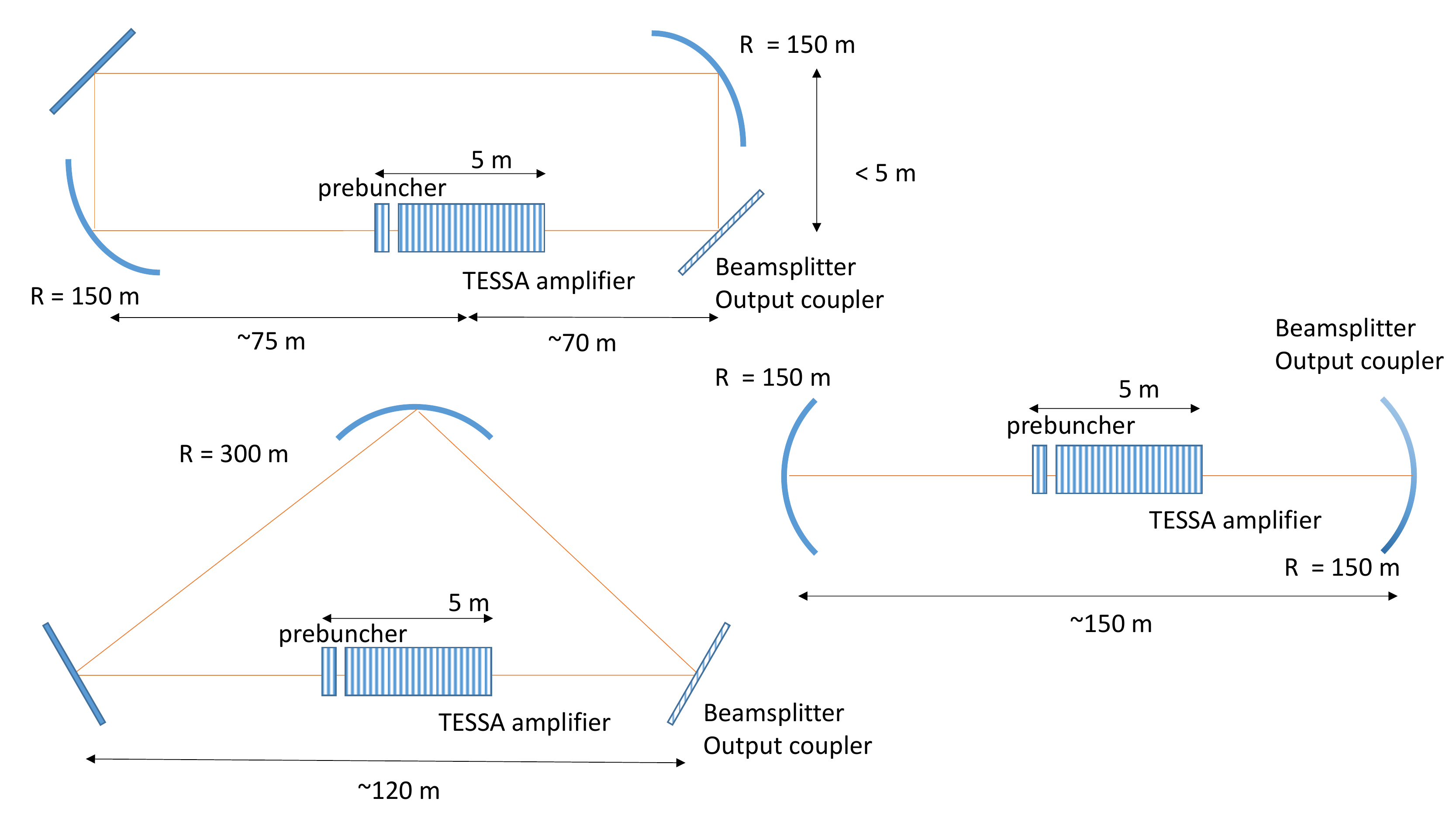}
\caption{Oscillator configuration. Near concentric resonator design. Design a) is a ring cavity and allows the injection of the igniter pulse. Design b) has less mirrors but presents some challenges for the injection. Design c) is a ring-like hemispheric resonator with only one curved mirror.}
\label{Fig:Setup}
\end{figure}

There are a variety of options for the ring cavity, including a three mirror design at the expenses of flexibility. In Fig. \ref{Fig:Setup} we show some of the cavity options and initial dimensions. While a full optical engineering design is not the scope of this paper, we only note here that cavities with such large optical powers have been in operation \cite{Carstens:2014}.

This simple design just provides a first order solution which is then refined using numerical methods. In practice, Genesis calculates the transverse radiation field profile evolution on a two-dimensional grid throughout the interaction within the undulator, and writes the output radiation field to disk. The radiation evolution along the recirculation cavity is modeled using the method of Huygens-Fresnel propagation on the output Genesis radiation profile (see Appendix A for details).

\section{Oscillator performance}

An oscillator simulation is performed by using the output of a single amplification stage as input for the next stage.  For a first pass at assessing the stability of the oscillator, we initially assume that the stimulated power simply adds to the seed laser mode. We can then use a given undulator taper to generate a map between input and output powers which can be used to study the dependence of the oscillator performance on the fraction of recirculated power.

\begin{figure}[t]
\includegraphics[width=125 mm]{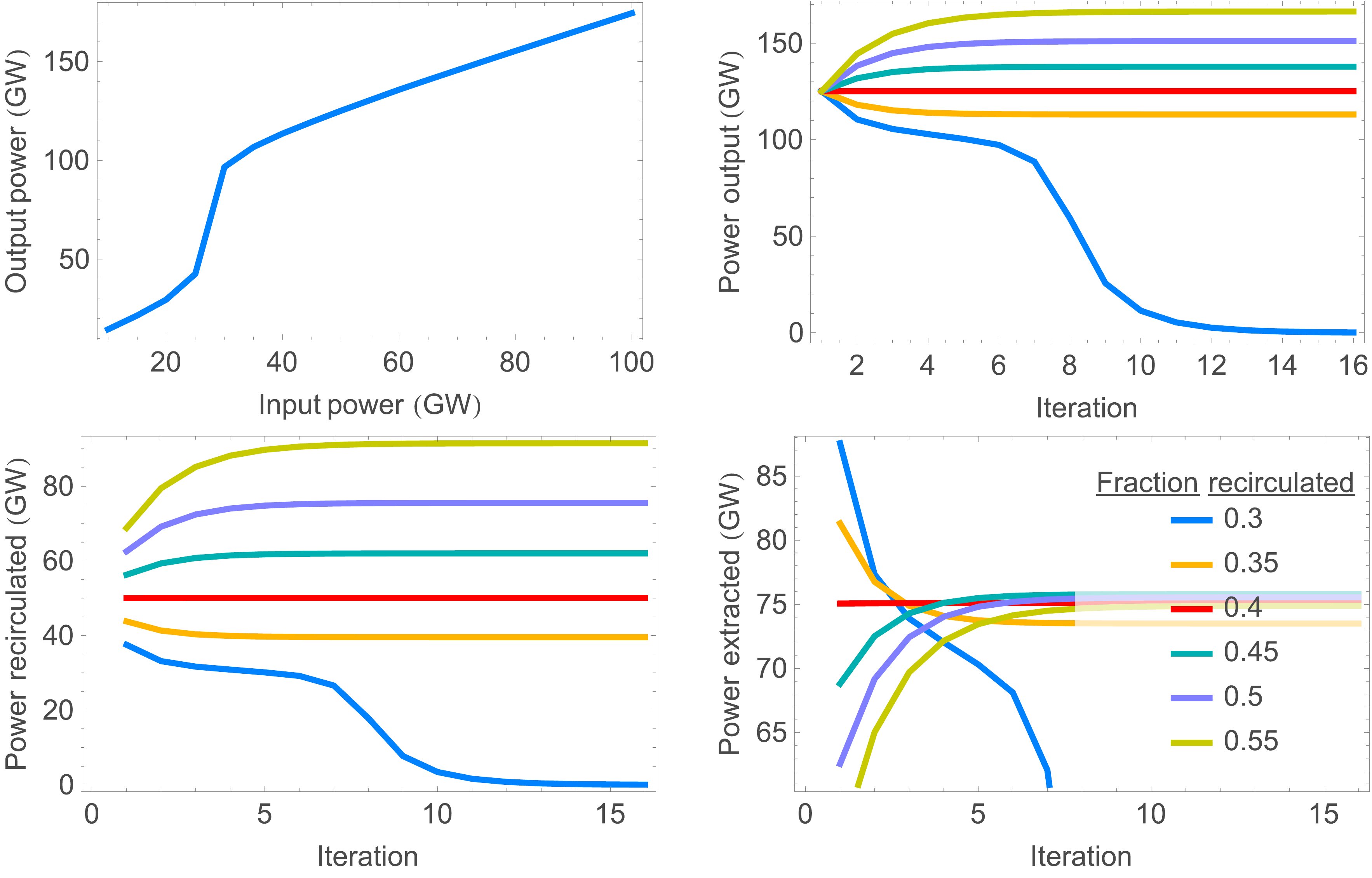}
\caption{Input-output power map (top left) is used to estimate oscillator performance for 50~GW startup power various power recirculation fractions.}
\label{Fig:PowerMap}
\end{figure}

Figure~\ref{Fig:PowerMap} shows the power map and power evolutions given a 50~GW startup seed power for various recirculation fractions. We find that at least 30$\%$ of the produced power must be fed into the undulator for the next pass to achieve stable operation.

% TODO: need to account for pulse stretching in these plots as well as cavity losses

As expected from Fig. \ref{Fig:PowerMap}, the stable solutions with different output couplers produce the similar power gain (about 75~GW) so that the additional power recirculated simply adds to the cavity power. Higher cavity powers may be detrimental as they increase the thermal loads on the cavity mirrors. On the other hand, higher cavity powers may make up for random perturbations in the e-beam parameters which pose a challenge in a real system.

Taking into account the 66\% output coupler reflectivity, diffraction and mirror losses of 20\%, and the slippage-induced losses of 33\%, the net fraction of power returned to the undulator is $\eta_s \eta_r \eta_l$ of the output power or 35\% in our example. The power converted from the beam energy per pass within the oscillator is 73~GW---nearly 150~mJ per pulse for the 2~ps long beam. Combined with the input power, the total power at the undulator's exit is 114~GW. The fraction of this power extracted from the cavity is $1-\eta_r=0.34$ so the net output power is 39~GW or 31\% of the beam power. Considering a 1~MHz electron beam repetition rate (250~kW average beam power), TESSO generates 78~kW of average output power with $\sim$31$\%$ efficiency.

\begin{figure}[t]
\includegraphics[width=160 mm]{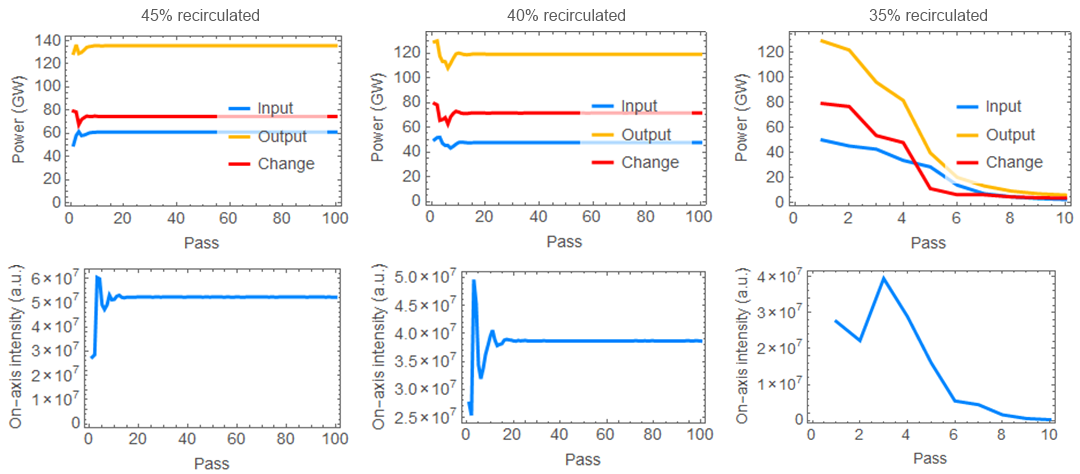}
\caption{Recirculation simulation using a full 3D model. The output Genesis field is recirculated via Huygens-Fresnel propagation.}
\label{Fig:RecircSim}
\end{figure}

When using the full three dimensional field to seed subsequent oscillator passes, we see (Fig.~\ref{Fig:RecircSim}) that the oscillator is stable for power recirculation fractions $\eta_r$ above 40\%, in agreement with our 1D estimate. Initially during the first few passes, transient fluctuations in the radiation size due to the competition between input and output radiation modes force the on-axis intensity to fluctuate before converging on a steady value. For lower power return fractions of 35\%, the power is so low that downward intensity fluctuations in the beam have a high probability of killing the oscillation. This fluctuation could be reduced by shaping the input mode, improving stability for lower power return fractions and recovering the previous result.

%Figure~\ref{Fig:ModeEvolution} shows the radiation mode evolution for the \oneum example with a curved mirror placed 71~m from the undulator's exit. The transverse distribution of the radiation quickly converges to a steady state profile over a few passes.
%
%\begin{figure}[t]
%\includegraphics[width=160 mm]{fig-mode_evolution}
%\caption{Radiation profile evolution from a Genesis simulation (bottom row) and after refocusing (top row). Each grid measures 9 mm by 9 mm and has $251^2$ cells.}
%\label{Fig:ModeEvolution}
%\end{figure}

Finally we should mention that in time-dependent simulations we observe the growth of the sideband instability, but the filtering in the spectral domain, which is used to stretch the pulse between each cavity pass, greatly suppresses this issue.

%__________________________________________________________________________________________________________
%__________________________________________________________________________________________________________
\section{Stability analysis}

%The stability and practicality of the double buncher arrangement should be considered.

%NICK has a figure about the stability of the double buncher for fluctuating input laser intensities, but we can keep that to answer referee's comments.
%Laser power fluctuations may perturb the first energy modulation amplitude, yet the mean energy of the beam should remain unchanged since the undulator is short and of constant period and strength. The chicane which follows delays the beam by the same amount regardless of the energy modulation, but the bunching of the detuned particles may vary depending on the energy modulation. The chicane's strength may be varied to inject the beam into the second energy modulator so that the mean energy does not change while the mean energy spread increases by the designed amount. If the mean energy of the beam has been preserved to this point, then the phase delay imparted by the final chicane should not vary.

We next consider the effects of random perturbations of the electron beam current on the oscillator stability. If the linac repetition rate is 3 orders of magnitude larger than that of the igniting laser, we should require stable operation for at least 1000 iterations on average.

The ponderomotive energy bandwidth is on the order of 0.5 MeV so shot-to-shot variations in the electron beam energy should be below 0.25 $\%$ which is well within the state-of-the-art especially for superconducting linac technology. On the other hand, shot-to-shot current variations may affect significantly the radiated power per pass which can reduce the recirculated power and stop the cavity lasing.

\begin{figure}[t]
\includegraphics[width=80 mm]{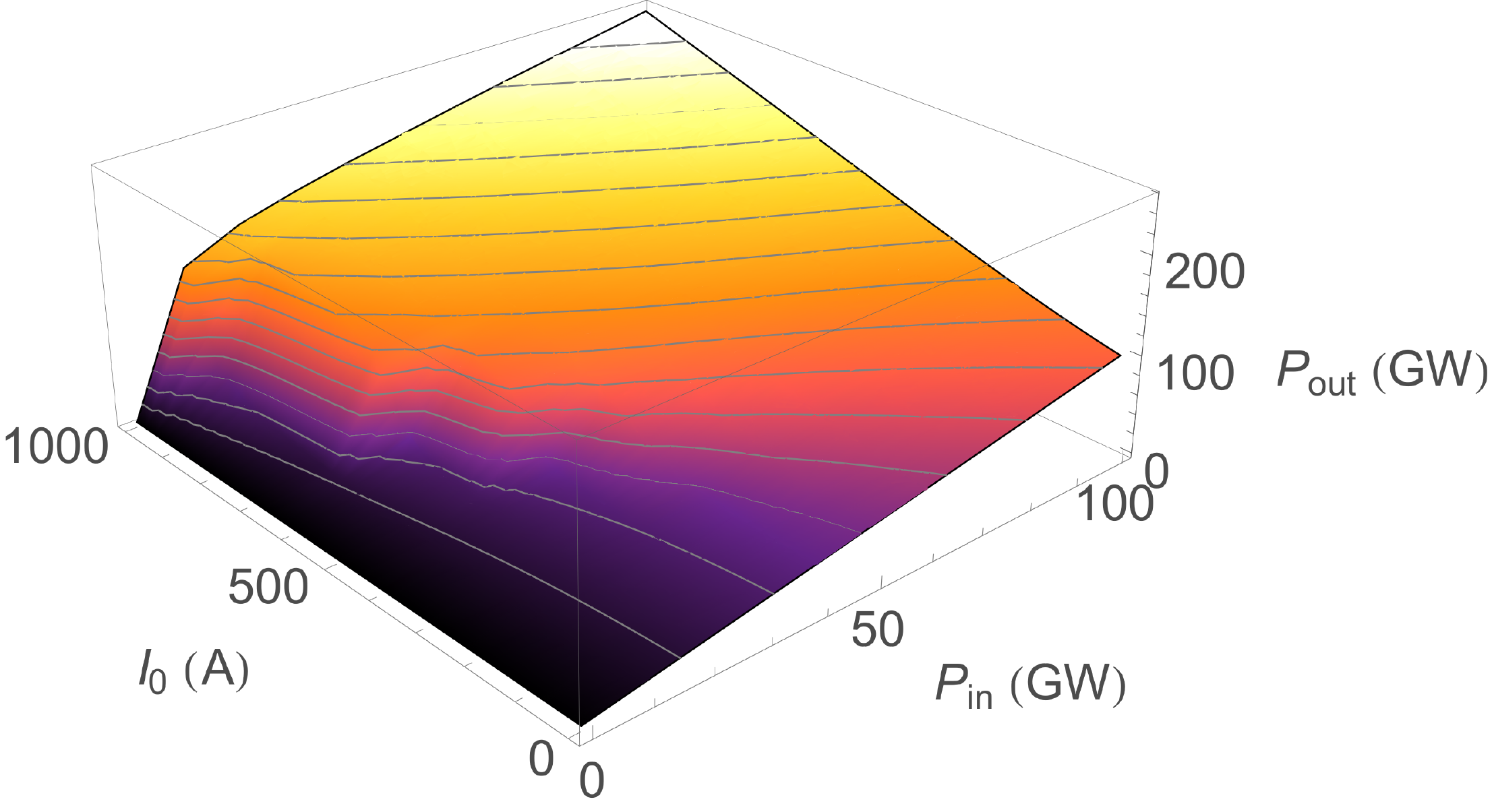}
\caption{Combined input power and current response map used to study the oscillator's resilience to current fluctuations.}
\label{Fig:PowerCurrentMap}
\end{figure}

In order to study the effect of these current fluctuations on the oscillator performance, we perform a Monte Carlo simulation using a power map accounting for various input currents in addition to input powers (shown in Figure~\ref{Fig:PowerCurrentMap}). Starting with an initial 50~GW startup seed power and a beam current randomly chosen from a normal distribution, we use the map to look up the output power and use a fixed fraction of that power and another randomly chosen current for the next iteration.

\begin{figure}[t]
\includegraphics[width=145 mm]{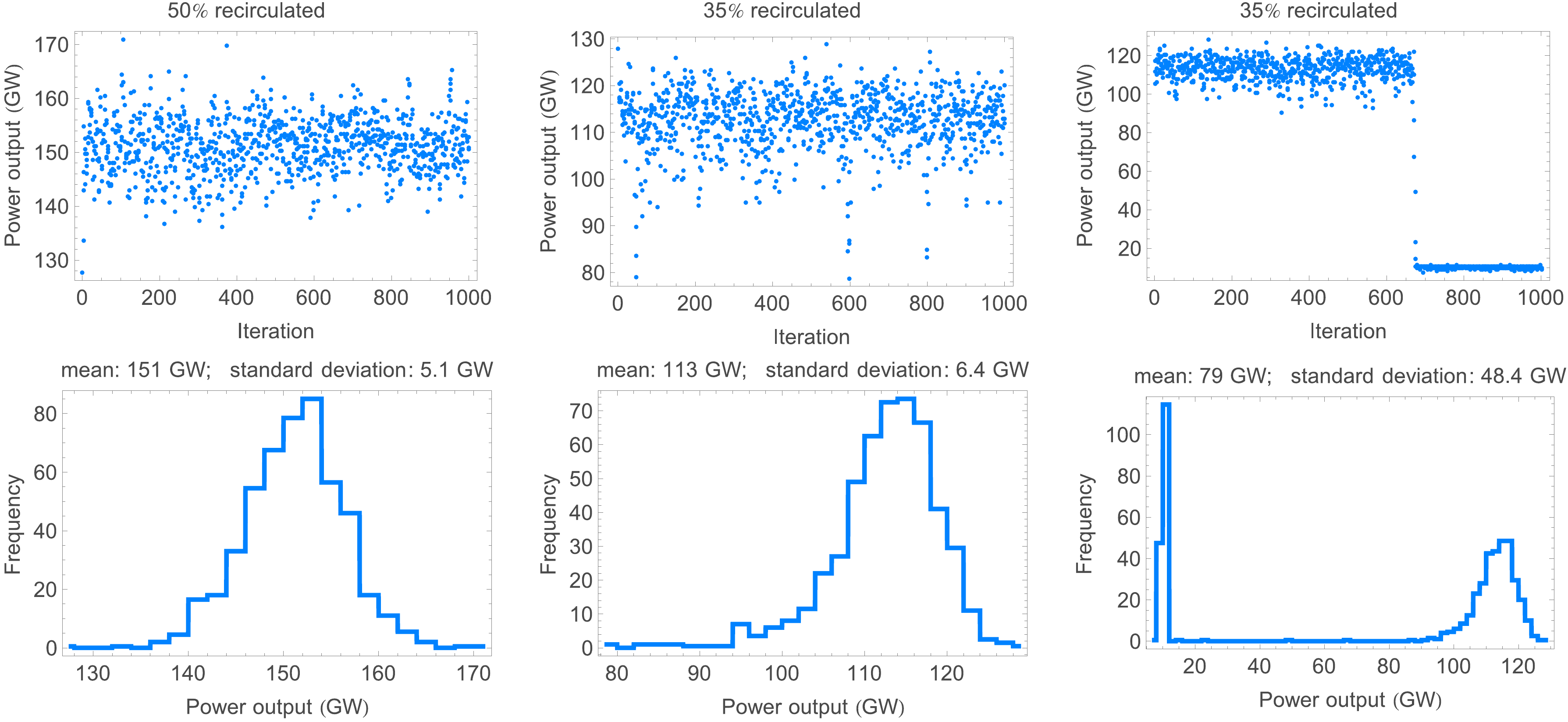}
\caption{Example power output Monte Carlo simulations for 5\%~rms input current variation.}
\label{Fig:CurrentVariationSingles}
\end{figure}

The result is a series of powers for each iteration which jitter about a mean value, and a few example results are shown in Figure~\ref{Fig:CurrentVariationSingles} for a constant initial power of 50~GW, 500~A average beam current with a 5\% rms variation, and recirculated power fractions of 35\% and 30\%. For 35\% returned power, the oscillator typically produces power stably for 1000 iterations and a representative shot is shown in the figure. As the fraction recirculated is decreased to 30\%, the produced power begins to fluctuate, decreasing the mean power relative to the median and increasing the variance. Occasionally, the power output fluctuates low enough for the 30\% recirculation case to stop power production until the next ignition pulse whereas this does not occur with 35\% recirculation which should then be considered when selecting an acceptable output coupler reflectivity.

\begin{figure}[t]
\includegraphics[width=120 mm]{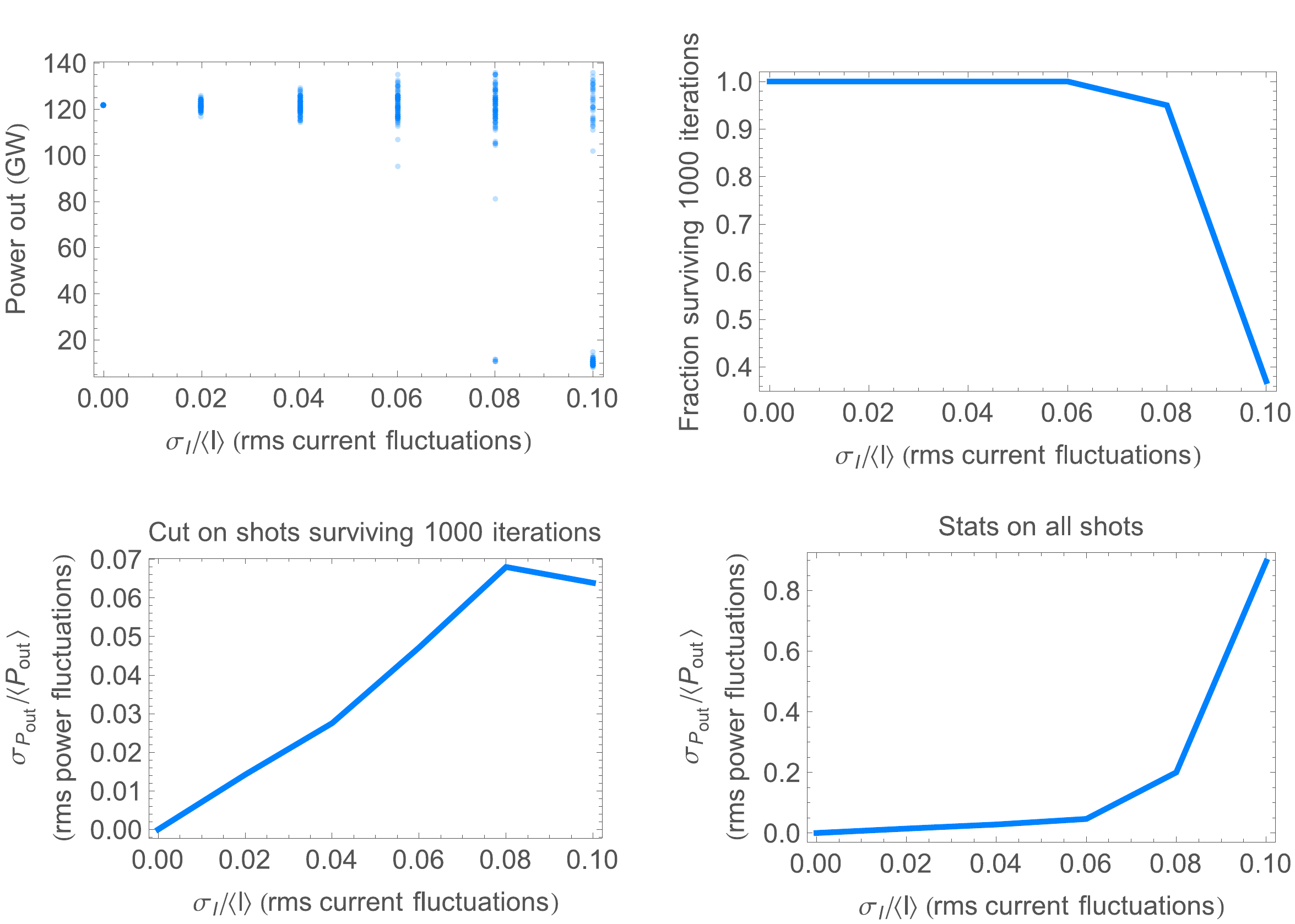}
\caption{Results of Monte Carlo simulations for 35\% return power fraction and various current fluctuations over 1000 iterations between subsequent ignitions. Top left: power output after 1000 iterations for various rms relative current fluctuations. Top right: fraction of ensemble surviving 1000 iterations. Bottom row: rms relative power variation after 1000 iterations for oscillations surviving 1000 iterations (left) and for all oscillations.}
\label{Fig:CurrentVariationStats}
\end{figure}

To quantitatively determine the maximum rms current fluctuations the oscillator can tolerate, we take the 35\% power return example and increase the relative rms current fluctuations until it degrades the oscillator performance. Figure~\ref{Fig:CurrentVariationStats} shows that the oscillator can survive up to 6\% rms relative current fluctuations. As current fluctuations are increased to 6\%, the rms relative power fluctuations increase proportionally; however, for current fluctuations greater than 6\%, a fraction of the oscillators stop lasing before the 1000th pass, degrading the average oscillator power. Nevertheless, better than 6\% current stability may be reliably achieved with linacs so this should not be a problem.

% for below plot: maybe delete lower left plot and maybe also perform stats on all passes * total population

Summarizing these stability requirements for a 500~A, 250~MeV electron beam, 4~m long undulator, and 50~GW seed, stable oscillator operation requires that we recirculate greater than 35 \% of the output power (assuming perfect TEM$_{00}$ input-output mode coupling), and keep the electron beam current fluctuations less than 6\% rms.

%__________________________________________________________________________________________________________
%__________________________________________________________________________________________________________
\section{Conclusions}

In conclusion, we have studied the possibility of embedding a TESSA amplifier in an optical cavity in order to reutilize a fraction of the output power as a seed for the high efficiency energy extraction. Fully self-consistent particle and field simulations show that $31~\%$ power extraction efficiency is possible with this system. In an example at 1 $\mu$m radiation wavelength, an average output power of $\sim$ 78 kW is obtained starting from an average beam power of 250 kW.

To simplify the discussion we have considered here the use of an igniter pulse. This is commercially available at 1 $\mu$m wavelength. When scaling at shorter wavelengths where a low repetition rate high power seed is not available, the oscillator start-up dynamics must be studied. In principle, using a dynamically controllable taper it will be possible to increase the energy extraction efficiency as the power builds up in the cavity until reaching the steady state, which would have the same characteristics as discussed here.

The results of this paper can be used as the basis of the design of a very high average power source which would enable many novel scientific and industrial applications.

\section{Acknowledgements}
This work was funded by DOE grant No. DE-SC0009914 and DOE SBIR grant No. DE-SC0013749. The authors wish to thank A. Zholents for very useful discussions. One of the authors (A. G.) is funded by a grant from the United States-Israel Binational Science Foundation(BSF), Jerusalem, ISRAEL.

\section{Appendix A}
The Huygens-Fresnel propagation method treats each point of radiation as a spherical wave which acquires a phase as it diffracts outward. For one transverse dimension, the spherical wave front, diffracting from a point over a distance $z_2-z_1=B$ through an optical element represented by an ABCD linear transport matrix, acquires a phase $-\frac{k}{2B} (A x_1^2-2 x_1 x_2+D x_2^2 )$ relative to an on-axis ray while the amplitude decreases to conserve power. Thus, the point spread function for one transverse dimension is given by
\begin{equation}
f(x_1,x_2)=\sqrt{i/B\lambda} e^{-i \frac{k}{2B}(A x_1^2-2 x_1 x_2+D x_2^2)}
\label{Eqn:PointSpreadFcn}
\end{equation}

Convolution of this point spread function with the initial radiation profile over the transverse grid containing the radiation yields the Huygens integral for one dimension
\begin{equation}
u_2(x_2)=\int_{-a_1}^{a_1} u_1(x_1)f(x_1,x_2) dx_1
\label{Eqn:HuygensIntegral}
\end{equation}

One limitation of this approach is that the calculated radiation must fit on the grid. While the radiation size at the entrance and exit of the undulator may be able to fit on the same size grid, at positions along the transport line such as at lenses where fluence considerations are important to study, the transverse beam size may be significantly larger. To overcome this limitation, we may scale the transverse coordinates at each position by different grid sizes by defining $\xi_1\equiv x_1/a_1$ and $\xi_2\equiv x_2/a_2=x_2/M a_1$ where $M\equiv a_2/a_1$ and then transform the input and output wavefunctions by \cite{Siegman}
\begin{eqnarray}
v_1(\xi_1)&\equiv& a_1^{1/2} u_1(x_1) e^{-i \frac{k}{2B} (a-M) x_1^2} \\
v_2(\xi_2)&\equiv& a_2^{1/2} u_2(x_2) e^{i \frac{k}{2B} (D-1/M) x_2^2}
\label{Eqn:CollimatingTransform}
\end{eqnarray}

With these scalings, the Huygens integral becomes
\begin{equation}
u_2(x_2)=\sqrt{i N_c} \int_{-1}^{1} v_1(\xi_1) e^{-i \pi N_c (\xi_1-\xi_2)^2} d\xi_1
\label{Eqn:HuygensIntegralTransformed}
\end{equation}
where $N_c \equiv a_1 a_2/B\lambda = M a_1^2 / B\lambda$ is the collimated Fresnel number. For a large number of grid points in two dimensions, numerical evaluation of this integral is computationally costly. In this case, numerical evaluation of the integral may be sped up significantly by using the convolution theorem to express the integral as the inverse Fourier transform of the product of the Fourier transforms of the input radiation profile and point spread function and employing a fast Fourier transform algorithm to evaluate the transforms. For this purpose, the spatial frequency projection of the point spread function is used to calculate the propagator
\begin{equation}
F(\kappa_x)=\frac{1}{\sqrt{2\pi}} e^{i \kappa_x^2 / 4 N_c \pi}
\label{Eqn:Propagator}
\end{equation}
where $\kappa_x$ is the spatial frequency conjugate to $\Delta\xi_x$. Since the largest value that $\Delta\xi_x$ takes is 1, we have $\kappa_x = 2\pi i$, where $i$ is an integer. Switching to 2D requires squaring the prefactors. Additionally, propagations may be chained with transverse amplitude or phase masks to simulate apertures and various focusing elements. Iteration of propagations and field dependent phase masks for small enough step sizes may even be used to model nonlinear effects in media (for example self-focusing of intense lasers in air).

\end{document}